\documentclass{PoS}
\usepackage{cite}

\title{Observations of supernova remnants and molecular clouds  from the mm to the gamma-ray domain:\\bridging low and high energy cosmic rays}

\ShortTitle{Bridging low and high energy cosmic rays}

\author{\speaker{Stefano Gabici}
\\
        APC, AstroParticule et Cosmologie, Universit\'e Paris Diderot, CNRS, CEA, Observatoire de Paris, Sorbonne Paris, 75205 Paris, France\\
        E-mail: \email{gabici@apc.in2p3.fr}}

\author{Thierry Montmerle\\
        Institut d'Astrophysique de Paris, 98bis Bd Arago, 75014 Paris, France\\
        E-mail: \email{montmerl@iap.fr}}

\abstract{
New evidence that cosmic rays (hadronic component) are accelerated by supernova remnant shocks all the way from low energies to high energies, has come from recent works combining gamma-ray observations in the sub-GeV to TeV domain on the one hand, and in the submm-mm domain on the other hand.  These observations concern the specific cases of supernova remnants interacting with molecular cloud complexes, that have long been suspected to be ideal laboratories to study in situ cosmic ray acceleration and diffusion. Indeed, enhanced gamma-ray emission from neutral pion decay, as well as enhanced ionisation (both by at least one order of magnitude with respect to average galactic values) have been observed in several regions of massive star formation housing supernova remnants interacting with molecular cloud complexes. This paper summarizes the main physical and chemical processes at work, as well as recent observations, that further support the paradigm of cosmic ray acceleration by supernova remnants all the way from the MeV domain up to several tens of TeV, although much work remains to be done to understand cosmic ray penetration and diffusion inside and around molecular clouds, and reveal the actual spectrum of the accelerated cosmic rays.
}

\FullConference{The 34th International Cosmic Ray Conference,\\
		30 July- 6 August, 2015\\
		The Hague, The Netherlands}

\begin{document}

\section{Historical overview and main scientific objectives}
\label{sec:intro}

In 1952, Hayakawa pointed out that cosmic rays\footnote{Here and in the following, with CRs we refer to the hadronic component of the cosmic radiation.} (CR) traversing interstellar matter would produce neutral pions that would in turn decay generating an observable flux of gamma rays \cite{hayakawa}. This early suggestion was confirmed in the late sixties by the detection by the OSO 3 satellite of a diffuse emission of $\gtrsim 100$~MeV photons from the Galactic plane \cite{oso3}. Such emission is naturally explained by the decay of neutral pions produced by CRs with a roughly homogeneous spatial distribution throughout the whole disk \cite{stecker}. 
However, the interstellar matter is far from being homogeneous, rather it is structured in diffuse gas, clumps, and clouds. It was soon realized, then, that due to their large masses, interstellar molecular clouds (MC) should appear as discrete gamma-ray sources on top of a ``truly'' diffuse Galactic emission \cite{blackfazio}.

The original proposal by Black \& Fazio \cite{blackfazio} was to infer the mass of a MC from its gamma-ray luminosity, and thus calibrate the quite uncertain CO to H$_2$ mass conversion factor $X_{\rm CO}$ (see \cite{lebrun} 
based on COS B data). This relies on the assumption of a uniform CR intensity throughout the Galaxy, which is certainly a reasonable one on large spatial scales, but very questionable on small ones. 
On the other hand, the argument can be reversed and,  after choosing a nominal value for $X_{\rm CO}$, the gamma-ray emission from a MC can be used to infer the CR intensity at the MC location \cite{issa}. 

However, some caveats are in order. First, the arguments described above do not hold if CRs do not penetrate efficiently MCs \cite{blitz} and, second, the quite large uncertainty of the measured value of $X_{\rm CO}$ induces an uncertainty of the same order into the estimate of the CR intensity based on gamma-ray observations, casting serious doubts on the feasibility of such approach \cite{hartquist}. 
The first issue was probably settled in \cite{skillingstrong}, 
where theoretical calculations were presented and it was concluded that, though an exclusion mechanism of CRs from MCs does exist, in most cases that would hardly work for proton energies larger than $\approx 100$~MeV. This value is safely below the threshold for neutral pion production in proton--proton interactions ($\sim 280$~MeV) and thus the gamma-ray emission from a MC should not, in general, be significantly affected by CR exclusion, a fact supported by GeV observations (see \cite{lebrunpaul} for early COS B results and \cite{fermipenetration} for recent FERMI observations).

The second issue is still open, and requires an assessment of the uncertainty of $X_{\rm CO}$. By combining the available observational measurements and the theoretical estimates of $X_{\rm CO}$ it is inferred that ``we know the mass-to-light calibration for giant MCs in the disk of the Milky Way to within $\pm 0.3$ dex certainly, and probably with an accuracy closer to $\pm 0.1$ dex (30\%). {\it This is an average number, valid over large scales}. Individual giant MCs will scatter around this value by a certain amount, and individual lines-of-sight will vary even more'' \cite{XCO}. Thus, we should probably accept as plausible an uncertainty of a factor of several in the estimate of the mass of MCs. Despite that, CR overdensities of an order of magnitude or more above the typical Galactic value would overwhelm the uncertainty on $X_{\rm CO}$ and thus would be revealed by means of gamma-ray observations of MCs.
Such large overdensities can be realised in the vicinity of a source of CRs \cite{atoyan}.

As early as 1934 Baade \& Zwicky \cite{zwicky} first pointed out that the energy released in supernova explosions suffices to explain the observed intensity of CRs\footnote{In fact, Baade \& Zwicky reached a correct result starting from erroneous assumptions. Indeed, the opening sentence of their paper reads: ``Two important facts support the view that CRs are of extragalactic origin''! See \cite{terhaar} for an early Galactic formulation of the supernova hypothesis.}, provided that a particle acceleration mechanism operates during or after the explosion and converts $\approx 10$\% of the explosion energy into CRs (i.e. each supernova should produce $\approx 10^{50}$ erg of CRs) . The expanding shocks driven by the supernova explosions in the interstellar medium, i.e. the supernova remnants (SNRs), were then identified as the most likely sites of production of Galactic CRs \cite{snrcr,ginzburg}, and the development of the theory of diffusive shock acceleration in the late seventies \cite{bobalsky} reinforced this hypothesis that, though very plausible, still remains to be proven (see e.g. \cite{hillas}). 

In 1979 one of us \cite{montmerle} discovered a spatial coincidence between SNRs, OB associations, and gamma-ray ``hot spots'' in COS B data. OB stars are massive and short lived, and thus do not have time to move far away from the site of their birth, i.e. a giant molecular cloud. 
Moreover, massive stars are the progenitors of core-collapse supernovae, and it is thus natural to expect a spatial correlation between OB associations, SNRs, and MCs.
It was immediately realized that the emission detected by COS B was too large to be explained by interactions of the uniform {\it sea} of Galactic CRs and the dense gas of the MCs: {\it an excess of roughly an order of magnitude in the flux of CRs was needed to match data}.
For this reason, the gamma-ray emission from such associations, dubbed SNOBs, was interpreted as the result of the decay of neutral pions produced in the interactions of CRs accelerated at SNRs with the dense gas of the parent MC. 
It follows that SNR/MC associations have the potential to offer a close-up view of the process of CR acceleration at shocks, and thus gamma-ray observations of such associations are of paramount importance in order to test the SNR hypothesis for the origin of Galactic CRs (see \cite{thierryreview} 
for recent reviews).

The poor angular resolution of COS B ($\approx 1^{\circ}$) represented the main limitation of these early studies, and as a consequence the associations of SNOBs with gamma-ray hot spots were only tentative (albeit convincing!). Despite tremendous improvements \cite{crushed}, the limited angular resolution remained a problem also for the GeV instruments that followed COS B, and the real breakthrough came with the advent of imaging atmospheric Cherenkov telescope arrays operating in the TeV energy domain which, thanks to the stereoscopic observation technique achieved a spatial resolution of the order of $0.1^{\circ}$ \cite{stereo}, needed to resolve MCs of size $\approx 10$~pc up to distances of several kpc.

\begin{figure}[t]
\centering
\includegraphics[width=0.55\textwidth,clip]{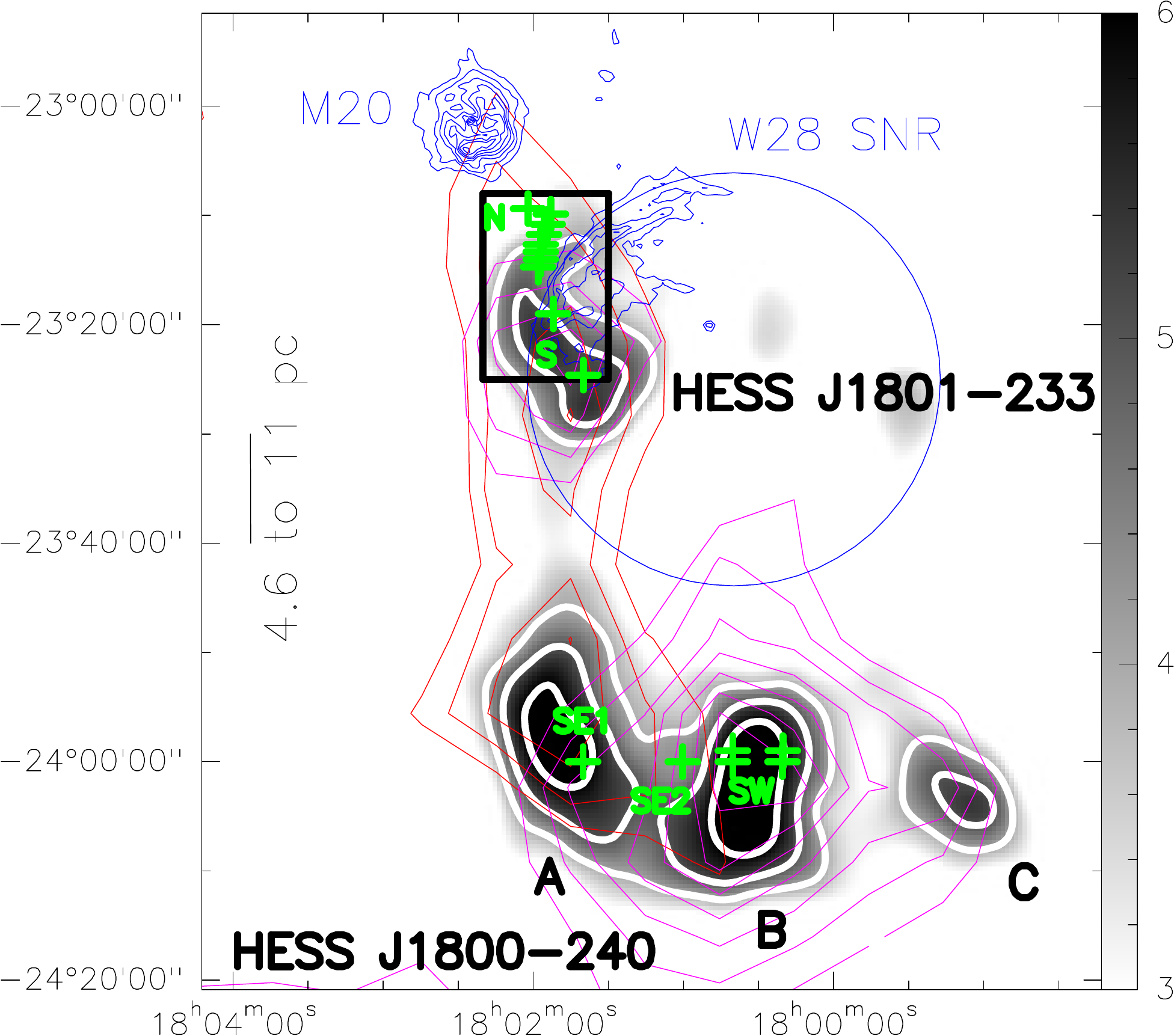}
\includegraphics[width=0.35\textwidth,clip]{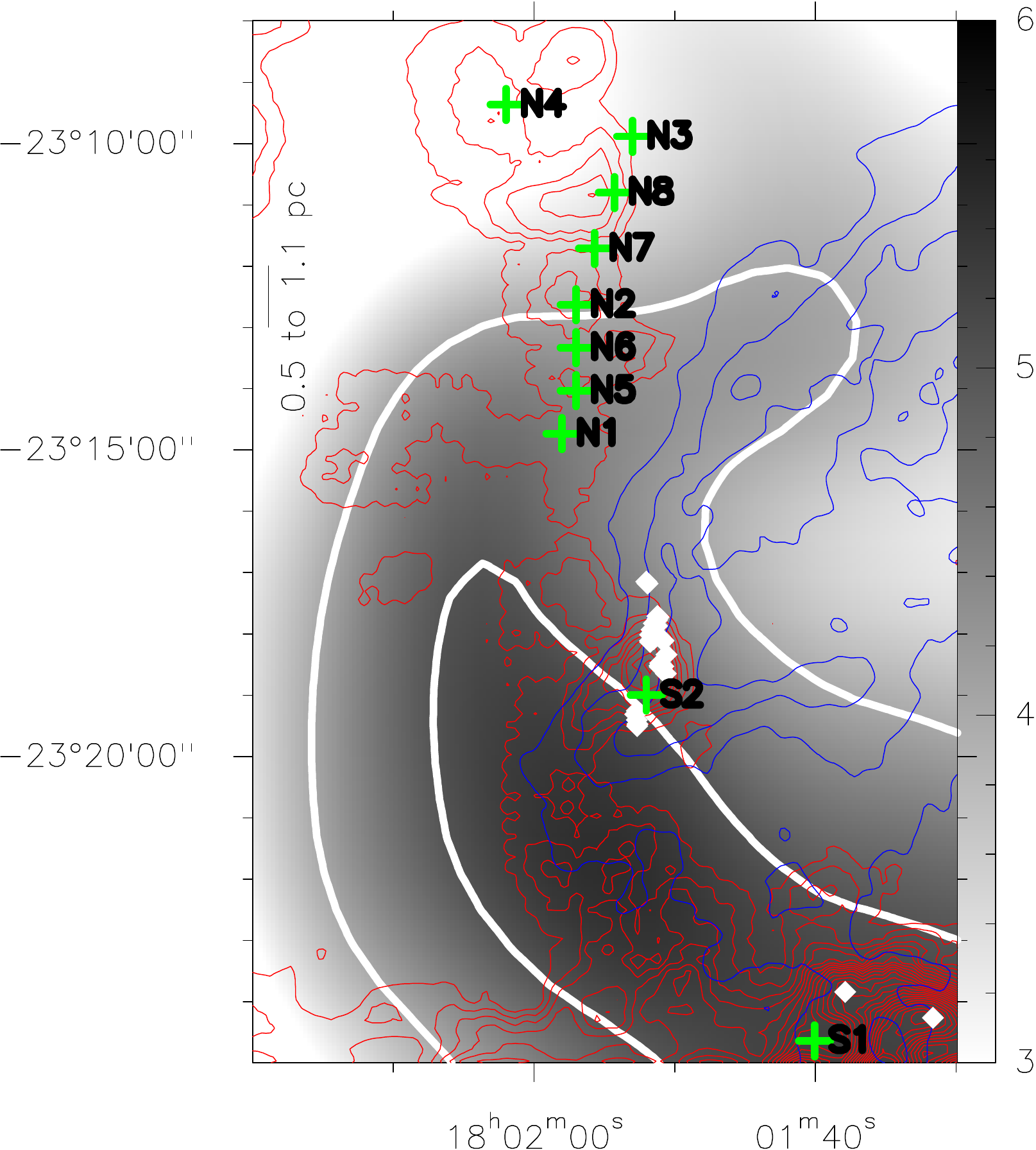}
\caption{{\bf Left:} Map of the W28 complex. Greyscale and white contours show the TeV emission seen by H.E.S.S. (levels are 4-6 $\sigma$), red (magenta) contours the CO(1-0) emission integrated over 15-25 km/s (5-15 km/s), and blue contours the 20 cm emission. The blue circle is the approximate boundary of the radio SNR. Green crosses correspond to mm (IRAM) observations aimed at determining the ionization rate of the molecular material close to the SNR shock. {\bf Right:} zoom on the black box \cite{w28noi}.}
\label{fig:w28}       
\end{figure}

A spectacular example of the impact of high resolution gamma-ray images is provided in Fig.~\ref{fig:w28}, where a map of the region surrounding the SNR W28 is shown.
The grey scale and white contours represent the TeV emission observed by H.E.S.S. \cite{hessw28}, which is found to correlate quite well with the distribution of molecular gas, as traced by the CO(1--2) emission (red and magenta contours, integrated over 15--25 and 5--15 km/s, respectively). The contour of the SNR radio shell is approximated by the blue circle. 
Notably, the gamma-ray emission from the southern part of the MC complex is clearly {\it not} overlapping with the SNR shell, but extends up to $\gtrsim 10$~pc ahead of the SNR shock.
The picture emerging from these observations is that of an association between a SNR and a MC complex emitting TeV gamma-rays (a SNOB!). 
Remarkably, the TeV emission detected from the MCs can be explained in terms of decay of neutral pions produced in CR interactions only if the intensity of multi-TeV CRs in the region exceeds by a factor of few tens the average Galactic one \cite{hessw28}.
This has been considered as an indirect evidence of the fact that the SNR W28 was, in the past, an efficient accelerator of particles of energy exceeding the TeV. Such particles escaped the SNR and now fill an extended region which encompasses the MCs which, being bombarded by CRs, appear as bright gamma-ray sources \cite{gabiciw28}. It follows that a good angular resolution is mandatory in order to identify the origin of the gamma-ray photons and, possibly, to distinguish the situation in which the gamma rays are produced at the site of interaction between a SNR shock and a MC , or well outside the SNR boundary (see e.g.\cite{hessw28,magicIC443,magicW51,fermiSNRs,fermiW44bis}).
From a theoretical point of view, the former scenario can be interpreted as the result of the acceleration or reacceleration of particles at shocks interacting with MCs \cite{blandfordcowie}, 
while the latter as the result of the interaction of runaway CRs that propagated from the parent SNR to a nearby MC \cite{atoyan,gabici09}. In both cases, an acceleration mechanism has to operate at SNR shocks.
We conclude, than, that {\it gamma-ray observations of SNR/MC associations are of paramount importance in order to understand the acceleration of particles at SNR shocks, their escape from the acceleration site, and their propagation in the interstellar medium. A full understanding of these issues is mandatory in order to finally verify (or falsify) the SNR paradigm for the origin of Galactic CRs.}

Given that the main goal of this paper is to describe an observational link between CRs of low and high energy, in order to proceed in our discussion we need an operational definition of {\it low} and {\it high} energy. 
A quite natural possibility is to choose as a dividing line the energy threshold for the production of neutral pions in proton-proton interactions, which is $E_{th} \sim 280$~MeV \cite{felixbook}. 
According to this definition, low energy CRs are those which do not produce neutral pions and thus cannot be traced by means of the gamma-ray observations described above. 
Even though CRs with energy below $E_{th}$ carry a subdominant fraction of the bulk energy of CRs (which is dominated by $\approx$~GeV particles), they nevertheless play a pivotal role in regulating the properties of the interstellar medium, by ionising and heating the gas, and thus driving interstellar chemistry in MCs, where ionizing photons cannot penetrate \cite{blackdalgarno,dalgarno}. 
Moreover, being the main regulators of the ionization fraction inside MCs, low energy CRs set the level of coupling between magnetic field and gas and thus influence the process of formation of stars and planets.
Hence, finding observational ways to trace them and quantify these effects is of prime interest.

Unfortunately, very little is known about the spectrum of low energy CRs in the interstellar medium, and even the determination of the local spectrum of such particles is problematic, due to the presence of the solar wind which prevents low energy particles to penetrate the solar system (an effect known as solar modulation \cite{solar modulation}). 
A way to quantify the impact of low energy CRs on the interstellar gas is by means of the ionization rate $\zeta_{\rm CR}$, which expresses the number of ionizations generated per second and per interstellar medium atom by CR interactions (see \cite{padovani} for a recent discussion), and is directly related to the CR heating rate of the interstellar gas \cite{fieldheating,glassgoldheating}.
Solar modulation affects the local CR spectrum below energies of several GeV, while CRs in the MeV domain are expected to dominate the ionization rate. 
For this reason, the first estimate of $\zeta_{\rm CR}$, performed in 1961, was based on a quite uncertain extrapolation of the observed local CR spectrum to low energies, and resulted in $\zeta_{\rm CR}^{\rm H} \sim 4 \times 10^{-16}$~s$^{-1}$ for an atomic hydrogen gas \cite{hayakawazeta}. 
The estimate was refined in 1968 by Spitzer \& Tomasko \cite{spitzer}, and then generalised to the case of a gas of molecular hydrogen in \cite{glassgoldH2}. This resulted in $\zeta_{\rm CR}^{\rm H_2} \sim 10^{-17}$~s$^{-1}$, which became a standard quantity known as the {\it Spitzer value}.
As noted in both \cite{hayakawazeta} and \cite{spitzer}, such estimate of $\zeta_{\rm CR}^{\rm H_2}$ had to be considered as a lower limit, being based on the extrapolation to low energies of the local CR spectrum, while much larger values were obtained (up to $\approx 10^{-15}$~s$^{-1}$) after assuming that the temperature of interstellar MCs is the result of the balance of CR heating and cooling processes.
However, the extrapolation down to 100 MeV of the local CR spectrum considered in \cite{hayakawazeta} and \cite{spitzer} turned out to be correct, as shown by recent measurement from the Voyager I probe, which reached the heliospheric boundary at a distance of 122 astronomical units from the Earth and probably measured the interstellar (i.e. unaffected by solar modulation) spectrum of CRs down to energies of a few MeV per nucleon \cite{voyager}, showing that a turnover is indeed present in the spectrum in the 100 MeV domain.

In fact, the discrepancy between the value of $\zeta_{\rm CR}^{\rm H_2}$ computed from the local CR spectrum and that inferred for MCs from thermal balance arguments is not necessarily a problem.
First of all, spatial variations of the intensity of low energy (significantly below 100 MeV) CRs are indeed expected throughout the Galactic disk, as a result of the short energy loss time of those particles that wouldn't allow them to reach large distances from their production sites and mix effectively to form a uniform background level \cite{cesarsky1975}. Thus, the local low energy CR spectrum is not necessarily representative of a typical Galactic value (as indeed it is for larger particle energies). 
Moreover, there is no {\it a priori} reason to believe the local CR spectrum to be representative of the spectrum of CRs inside MCs, but rather the contrary.
This is because severe energy losses (mainly ionization losses) in dense MCs would cool the CR spectrum, and replenish with particles the energy region below $\approx 100$ MeV.
Thus, a better understanding of the mechanism of CR penetration into MCs is needed in order to obtain realistic estimates of $\zeta_{\rm CR}^{\rm H_2} $.
Finally, to complete (and complicate) the picture, the relative weight of the CR proton and electron contribution to the ionization rate is, to date, not at all clear \cite{padovani}, even though the recent Voyager I measurements of the local CR spectrum (of both protons and electrons) might shed new light on this issue.

Observationally, there is a powerful way to measure the ionisation rate in MCs 
based on the detection of molecular lines in MC spectra. 
The discovery, in 1970, of a new molecular emission line in the millimetre domain (89.188 GHz) \cite{unidentifiedline} associated to the HCO$^+$ molecular ion \cite{HCO+}, was accompanied by a discussion on the importance of CRs on the chemistry of diffuse and dense clouds \cite{herbst}.
Indeed, the abundance of HCO$^+$ is connected to the CR ionization rate because the abundances of this specie depends on a quite small number of chemical reactions, initiated by the CR ionization of H$_2$ which leads to the formation of the protonated molecular hydrogen H$_3^+$.
HCO$^+$ is then produced in dense clouds as the result of the reaction of H$_3^+$ with CO.

It follows that a straightforward way to measure the CR ionisation rate is to search for lines associated to the pivotal H$_3^+$ ion \cite{oka}. These are absorption lines in the infrared domain.
Due to the extremely simple chemistry involved in the formation of H$_3^+$, its abundance is directly connected to the CR ionization rate \cite{dalgarno,oka}.
H$_3^+$ was first detected in interstellar space in 1996 \cite{geballe}, and shortly after its detection was reported from several dense MCs, roughly at the level predicted by chemical models after assuming a ``canonical'' value for $\zeta_{\rm CR}^{\rm H_2}$ \cite{mccall1999}. 
On the other hand, contrary to expectations, large abundances of H$_3^+$ were also found in diffuse clouds \cite{geballe1999,mccall1998}. 
However, theoretical predictions of the H$_3^+$ abundance suffered from large uncertainty on several key parameters, so this finding was initially taken with great caution. In particular, measured values of the H$_3^+$ destruction rate by electrons (dissociative recombination, the main destruction channel of H$_3^+$ in diffuse clouds) from different techniques were found to disagree by several orders of magnitude\cite{larsson}! The laboratory measurement of the H$_3^+$ destruction rate under nearly interstellar condition was then a major step forward, and gave a value of $2.6 \times 10^{-7}$~cm$^3$/s at a temperature of 23 K \cite{mccallnature}. This allowed a reliable estimate of the abundance of H$_3^+$ in diffuse clouds and called for an enhanced (with respect to the Spitzer value) CR ionisation rate at the level of $\zeta_{\rm CR}^{\rm H_2} \approx 10^{-15}$~s$^{-1}$.

Observations of lines from HCO$^+$ and from its isotopologue DCO$^+$ from dense MCs also allow to determine the CR ionization rate.
In a pioneer paper, Gu\'elin et al. (1977) \cite{guelin} suggested to use the abundance ratio of DCO$^+$ (first detected in 1976 \cite{DCO+}) and HCO$^+$, $R_{\rm D} = n({\rm DCO}^+)/n({\rm HCO}^+)$, to measure the ionization fraction $x_e$ in MCs.
It was then proposed to use $R_{\rm D}$ in combination with the abundance ratio $R_{\rm H} = n({\rm HCO^+})/n({\rm CO})$ to derive both $x_e$ and the CR ionisation rate $\zeta_{\rm CR}^{\rm H_2}$ in dense MCs \cite{caselli}.
Measurements of the $R_{\rm D}$ and $R_{\rm H}$ ratios from a number of dense clouds resulted in a CR ionization rate roughly at the level of the Spitzer value \cite{caselli}.

The implicit assumption here  is that UV photons, the main ionizing agents of the interstellar medium, do not play a role in ionizing MCs, being absorbed by a thin layer of gas corresponding to $\sim 4$ magnitudes of visual extinction \cite{mckeeUV}. It is only under these circumstances that molecular abundances are related to the CR ionization rate only, and not to the photoionization rate. 

\begin{figure}[t]
\centering
\includegraphics[width=0.5\textwidth,clip]{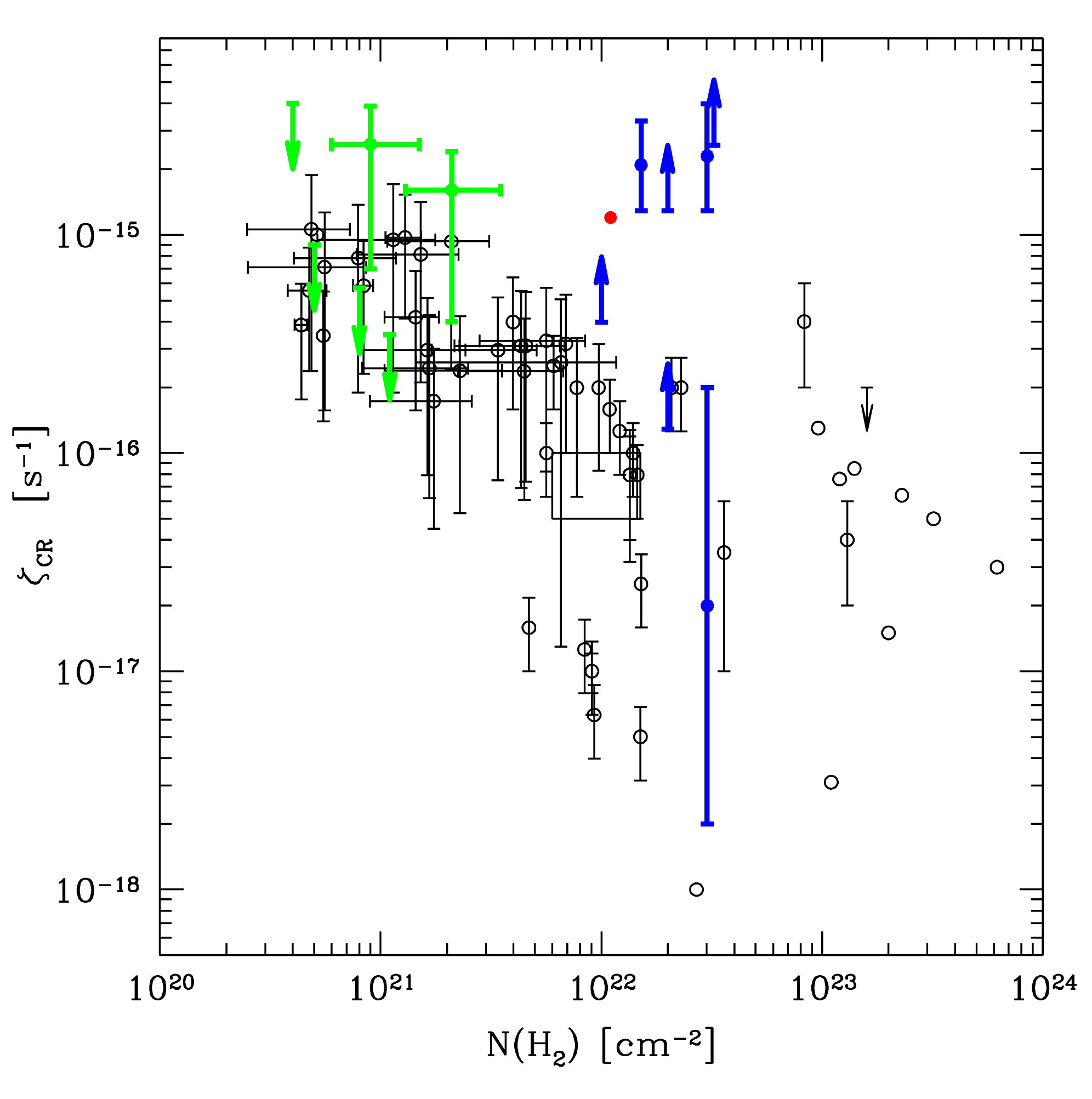}
\caption{CR ionization rate versus MC column density. Black data points refer to isolated MCs \cite{marcoreview}, while colored ones to the SNR/MC associations IC443, W51C, and W28 (green, red, and blue points \cite{indrioloIC443,ceccaW51c,w28noi}).}
\label{fig:zeta}       
\end{figure}

A compilation of the measurements of $\zeta_{\rm CR}^{\rm H_2}$ in MCs, obtained mainly by means of the two approaches described above, is given in Fig.~\ref{fig:zeta}. 
Black data points refer to isolated (and gamma-ray quiet) MCs, while colored ones to gamma-ray-bright SNR/MC associations, i.e. MCs spatially associated with SNRs. 
The presence of gamma-ray emission indicates that the MC material is bombarded by an enhanced (with respect to the CR background) flux of high energy (GeV/TeV) CRs. 
It is clear from Fig.~\ref{fig:zeta}  that a trend exists for isolated MCs, namely, the CR ionization rates  are a factor of $10 ... 100$ larger in diffuse clouds (column densities $N_{\rm H} \approx 10^{21} ... 10^{22}$~cm$^{-2}$) than in dense clouds ($N_{\rm H} > 10^{22}$~cm$^{-2}$), where it is roughly comparable to the Spitzer value. On the contrary, a different trend is observed for MCs located next to gamma-ray-bright SNRs, for in this case the ionization rates reach large values of the order of $\gtrsim 10^{-15}$~s$^{-1}$, regardless of the cloud column density. 
Even though this result should be taken with care, being based on the observation of three SNR/MC associations only, it should be nevertheless considered as a first evidence for the fact that MCs belonging to a gamma-ray bright SNR/MC association exhibit, as a general behavior, larger values of the CR ionization rate when compared to isolated clouds. 
In other words, such MCs are bombarded by a larger flux of low energy (probably $\sim$MeV), ionizing CRs.
The associated SNR shock is the most plausible site for the acceleration of suck particles.
If confirmed by further observations, this scenario would fit with the popular (but still not proven) idea according to which SNRs accelerate Galactic CRs. 
It is a remarkable fact that {\it the combination of low and high energy observations of SNR/MC associations, and the emerging evidence of a correlation between large gamma-ray fluxes and enhanced ionization rates provide not only additional support to the SNR hypothesis for the origin of Galactic CRs, but also establishes a long sought connection between low and high energy CRs.} 
This will allow us to test models on CR acceleration and propagation over an energy interval of unprecedented breadth, spanning from the MeV to the TeV domain.

The paper is organized as follows: in Sec. \ref{sec:gamma} and \ref{sec:chemistry} we discuss the physics of CR interactions in MCs and how they determine both the gamma-ray emission and the chemistry in MCs.
In Sec. \ref{sec:SNRMC} we review high and low energy observations of SNR/MC associations. 
We conclude in Sec. \ref{sec:conclusions}.

\section{Gamma rays from dense molecular clouds}
\label{sec:gamma}

In this Section we develop a simplified formalism to compute the gamma-ray emission from a MC of a given mass bombarded by CRs of arbitrary intensity. For simplicity we limit ourselves to the case of CR spectra which follow power laws in particle energy.

Consider a MC of mass $M_{cl}$ pervaded by a spatially uniform distribution of CRs (protons) with power law spectrum $n_{\rm CR}(E) \propto E^{-\alpha}$ with $\alpha > 2$ [cm$^{-3}$ TeV$^{-1}$] defined as:
\begin{equation}
\label{eq:CRspectrum}
E^2 n_{\rm CR}(E) = (\alpha - 2) \delta ~ w_{\rm CR}^0(> 10~{\rm TeV}) \left( \frac{E}{10~{\rm TeV}} \right)^{2-\alpha}
\end{equation}
where the normalization has been chosen in such a way that the energy density of CRs above particle energy 10~TeV is $\delta$ times that of the Galactic CR sea that pervades the whole Galactic disk, $w_{\rm CR}^0(>10~{\rm TeV}) \approx 10^{-3}$~eV/cm$^3$. In other words, $\delta$ represents the overdensity of multi-TeV CRs at the MC location, and is likely to exceed significantly unity if the MC is located in the vicinity of a CR accelerator (i.e. a SNR). 
Gamma rays are then produced as the result of proton-proton interactions of CRs with the intercloud gas of hydrogen density $n_{cl}$ through the reaction chain:
\begin{equation}
p + p \rightarrow p + p + \pi^0  ~,~~~~~~~~~~
\pi^0 \rightarrow \gamma + \gamma ~ . 
\end{equation}
The first reaction -- the neutral pion production -- happens if the kinetic energy of the CR proton is above a threshold of $E_{th} \approx 280$~MeV \cite{felixbook}.
For energies much larger than $E_{th}$ the resulting gamma-ray emissivity is roughly of the order of:
\begin{equation}
E_{\gamma}^2 q_{\gamma}(E_{\gamma}) \approx \eta_N \frac{E^2 n_{\rm CR}(E)}{\tau_{pp}^{\pi^0}} 
\end{equation}
where $E_{\gamma} \approx E/10$ is the energy of the gamma ray as a function of the parent CR proton energy, $\tau_{pp}^{\pi^0} = 1.6 \times 10^8 (n_{cl}/{\rm cm}^3)^{-1}$~yr is the energy loss time of CRs due to $\pi^0$ production, and $\eta_N \approx 1.8$ is an enhancement factor to account for the presence of nuclei heavier than hydrogen in both CRs and interstellar gas \cite{felixbook}.
By combining the equations above one gets a gamma-ray flux of:
\begin{equation}
\label{eq:gammarayflux}
F_{\gamma}(E_{\gamma}) \approx 3.6 \times 10^{-13} ~\delta (\alpha - 2) \left( \frac{M_{cl}}{10^5 M_{\odot}} \right) \left( \frac{d}{\rm kpc} \right)^{-2} \left( \frac{E_{\gamma}}{\rm TeV} \right)^{-\alpha} {\rm TeV^{-1} cm^{-2} s^{-1}}
\end{equation}
or, equivalently, an integral gamma-ray flux of:
\begin{equation}
\label{eq:integral}
F_{\gamma}(E_{\gamma} > 1 ~ {\rm TeV}) \approx  3.6 \times 10^{-13} ~\delta ~ \frac{\alpha - 2}{\alpha - 1} \left( \frac{M_{cl}}{10^5 M_{\odot}} \right) \left( \frac{d}{\rm kpc} \right)^{-2} {\rm cm^{-2} s^{-1}}
\end{equation}
for a cloud at a distance $d$. Though approximate, Eqns.~\ref{eq:gammarayflux} and \ref{eq:integral} are in reasonable agreement with more detailed calculations \cite{felixclouds}.

One sees from Eq.~\ref{eq:gammarayflux} that MCs can be used as \textit{CR barometers}: if the mass and distance of the cloud is known, the CR overdensity $\delta$ can be deduced by the observation of the cloud gamma-ray spectrum \cite{felixclouds,sabrinona}.
However, this can be done only as long as the cloud is detectable by a given telescope.
The point source sensitivity of current generation Cherenkov telescope arrays at photon energies above 1 TeV is of the order of $\Phi_{\gamma}(> 1~{\rm TeV}) \approx 10^{-12}$~cm$^{-2}$ s$^{-1}$, and it worsens for extended sources as $\approx \vartheta_{cl}/\vartheta_{res}$, where $\vartheta_{cl}$ is the source size and $\vartheta_{res}$ their typical angular resolution of the order of a tenth of a degree.
Given the fact that MCs are extended objects of apparent size:
\begin{equation}
\vartheta_{cl} \approx 1^{\circ} \left(\frac{M_{cl}}{10^5 M_{\odot}}\right)^{1/3} \left( \frac{n_{cl}}{100~{\rm cm}^{-3}} \right)^{-1} \left( \frac{d}{1~{\rm kpc}} \right)^{-1}
\end{equation}
it follows that, unless the MC is extremely massive and/or extremely nearby, the condition of detectability is $\delta \gg 1$, i.e. large CR overdensities are required to detect TeV gamma rays from MCs.
Thus, the gamma-ray bright MCs belonging to the SNR/MC associations in Fig.~\ref{fig:zeta} (colored points) are bombarded by CRs of intensity much larger than the typical one in the Galactic disk.

\section{Cloud chemistry, molecular lines, and the cosmic ray ionization rate}
\label{sec:chemistry}

In this Section we review the chemistry driven by CRs in MCs. For sufficiently large column densities (corresponding to $\sim 4$ magnitudes of visual extinction) CRs are the only ionizing agents able to penetrate MCs. In particular, CRs ionize molecular hydrogen which in turn produces, via an ion-neutral reaction, the pivotal ion H$_3^+$, which plays a central role in the MC chemistry \cite{oka,dalgarno}.

H$_3^+$ is generated in MCs via the reaction chain:
\begin{equation}
\label{eq:H3+}
{\rm CR} + {\rm H}_2 \longrightarrow {\rm CR} + {\rm H}_2^+ + e^- ~, ~
{\rm H}_2^+ + {\rm H}_2 \longrightarrow {\rm H}_3^+ + {\rm H}
\end{equation}
where CRs are assumed to be the unique responsible for the ionization of H$_2$.
The Langevin constant for the second reaction in (\ref{eq:H3+}) is $k_L \sim 2 \times 10^{-9}$~cm$^3$/s and thus the reaction rate $k_L n({\rm H}_2)$ is orders of magnitude larger than the CR ionization rate $\zeta_{\rm CR}^{\rm H_2}$ under any plausible assumption. This implies that the production rate of H$_3^+$ per unit volume is simply given by $\zeta_{\rm CR}^{\rm H_2} n({\rm H}_2)$.

In diffuse clouds H$_3^+$ is destroyed by recombination with electrons released by the photoionization of C into C$^+$ (the most abundant ion in diffuse MCs), at a rate $k_e n({\rm H}_3^+) n(e^-)$, with $k_e \approx 2.6 \times 10^{-7}$~cm$^3$/s (at $T = 23$~K \cite{mccallnature}). The balance of production and destruction rates gives:
\begin{equation}
\zeta_{\rm CR}^{\rm H_2} ~=~ k_e ~ n({\rm H}_3^+) ~ \frac{n(e^-)}{n({\rm H}_2)} ~=~ \frac{2 ~k_e~ x_e}{f_{{\rm H}_2}} ~ n({\rm H}_3^+) ~\sim~ \frac{2 ~k_e~ x_e}{f_{{\rm H}_2}} ~ \frac{N({\rm H}_3^+)}{L}
\end{equation}
where we introduced the electron fraction $x_e = n(e^-)/n_{\rm H}$ and the fraction of hydrogen nuclei in molecular form $f_{{\rm H}_2} = 2 n({\rm H}_2)/n_{\rm H}$, with $n_{\rm H} = 2 n({\rm H}_2) + n({\rm H})$. In the last (approximate) equality we assumed spatial homogeneity and passed from volume densities $n$ to column densities $N = n/L$, $L$ being the cloud depth. For reference values of diffuse MCs $x_e = x({\rm C}^+) \sim 1.5 \times 10^{-4}$ and $0.67 < f_{{\rm H}_2} < 1$ (0.67 corresponds to equal number of H and H$_2$) \cite{indriolo2012} one gets \cite{oka}:
\begin{equation}
\label{eq:zetadiffuse}
\zeta_{\rm CR}^{\rm H_2} ~\sim~ \frac{2.6 \times 10^{-15}}{f_{{\rm H}_2}} \left[ \frac{N({\rm H}_3^+)}{10^{14}~{\rm cm}^{-2}} \right] \left( \frac{L}{\rm pc} \right)^{-1} ~ {\rm s}^{-1}
\end{equation}
It follows that, if $L$ is estimated from the cloud angular size after assuming rough spherical symmetry, the CR ionization rate can be estimated form the measurement of the H$_3^+$ column density. 

Under typical MC conditions only the two lowest levels of H$_3^+$ are significantly populated (see Fig.~1 from \cite{mccall1999}), and luckily a number of spectral lines originating from the transitions from these two levels fall in a region of the infrared spectrum observable by ground based telescopes, e.g. the $R(1,1)^u$ and $R(1,0)$ transitions are observable at a wavelength of 3.668083 $\mu$m and 3.668516 $\mu$m, respectively. It has to be stressed that the H$_3^+$ lines are detected in {\it absorption}, and thus an estimate of $N({\rm H}_3^+)$ from a given MC can be done solely in the presence of a bright infrared foreground (or embedded) star. Indriolo and McCall \cite{indriolo2012} presented a survey of H$_3^+$ towards isolated diffuse MCs (no known gamma-ray-bright SNRs in their vicinity). Amongst the 50 lines of sight examined, 21 resulted in a detection of H$_3^+$. The mean value of the CR ionization rate derived in this way was $\zeta_{\rm CR}^{{\rm H}_2} \sim 3.5 \times 10^{-16}$ s$^{-1}$, a factor of several tens larger than the Spitzer value.
Moreover, as long as only diffuse MCs are considered, no dependence of the CR ionization rate on the cloud column density has been found.
In other words, the mean value given above can be taken as a typical reference value for the CR ionization rate in diffuse clouds.

The situation changes dramatically if we consider larger column densities, i.e. dense clouds. There, carbon atoms are mostly in {\rm CO}, and H$_3^+$ is mainly destroyed by the reaction:
\begin{equation}
\label{eq:HCO+}
{\rm H}_3^+ + {\rm CO} \longrightarrow {\rm HCO}^+ + {\rm H}_2
\end{equation}
characterized by a rate $k_{\rm CO} \sim 2 \times 10^{-9}$~cm$^3$/s at typical MC temperatures.
In this context, an expression similar to Eq.~\ref{eq:zetadiffuse} can be derived for dense clouds:
\begin{equation}
\zeta_{\rm CR}^{\rm H_2} ~\sim~ 2 \times 10^{-17} \left[ \frac{N({\rm H}_3^+)}{10^{14}~{\rm cm}^{-2}} \right] \left( \frac{L}{\rm pc} \right)^{-1} ~ {\rm s}^{-1}
\end{equation} 
indicating that for $N({\rm H}_3^+)$ of the order of several $10^{14}$~cm$^{-2}$ the CR ionization rate is quite close to the Spitzer value. Such H$_3^+$ column densities have indeed been observed from a number of dense MCs \cite{geballe,mccall1999}.
Thus, while the value of $\zeta_{\rm CR}^{\rm H_2}$ found in diffuse MCs does not depend on the hydrogen column density, it does when the transition between the diffuse to dense MC regime is probed \cite{indriolo2012}.

A difficulty with the detection of H$_3^+$ from dense clouds is the fact that for large gas column densities the background infrared source might be totally obscured by the cloud itself. Thus, alternative approaches to estimate $\zeta_{\rm CR}^{\rm H_2}$ have been sought. A natural possibility is to search for molecular lines associated to the molecular ions HCO$^+$ and DCO$^+$ (see \cite{paola} and references therein for a detailed discussion). The advantage of this approach resides in the fact that such lines are in emission (no need for a background source!) and fall in the millimeter domain, which is probed by instruments like the 30 meters IRAM telescope. Another advantage of emission lines is that they allow a detailed mapping of extended clouds, a thing which is unfeasible by means of observations of absorption lines, since one would need an unreasonably large number of background sources.

Similarly to HCO$^+$, DCO$^+$ is mainly produced by the reaction:
\begin{equation}
\label{eq:DCO+}
{\rm H}_2{\rm D}^+ + {\rm CO} \longrightarrow {\rm DCO}^+ + {\rm H}_2
\end{equation}
The reaction rates of H$_2$D$^+$ and H$_3^+$ with CO are similar, as well as the destruction rates of HCO$^+$ and DCO$^+$ via dissociative recombination. From this, and from the fact that the competing reaction H$_2$D$^+$+CO $\rightarrow$ HCO$^+$+HD proceeds with a branching ration twice that of reaction \ref{eq:DCO+} one can easily infer that $R_{\rm D} = n({\rm DCO}^+)/n({\rm HCO}^+) \approx n({\rm H}_2{\rm D}^+)/3 n({\rm H}_3^+)$.

The abundance ratio of H$_2$D$^+$ over H$_3^+$ can be then derived by recalling that the former molecule is produced in the exothermic reaction:
\begin{equation} 
{\rm H}_3^+ + {\rm HD} \longrightarrow {\rm H}_2{\rm D}^+ + {\rm H}_2
\end{equation} 
characterized by a reaction rate $k_{\rm D}$. The reverse reaction is endothermic with an energy barrier of $\sim 220$ K, and it is thus very inefficient at the low temperatures typical of MCs.
This implies that the enhancement in the abundance ratio of H$_2$D$^+$ over H$_3^+$ in clouds is limited by the rate at which electrons and neutrals (mainly CO) destroy H$_2$D$^+$, with a reaction rate of $k_e^{\prime}$ and $k_{\rm CO}$, respectively. 
After some manipulations this leads to \cite{paola}:
\begin{equation}
R_D = \frac{n({\rm DCO}^+)}{n({\rm HCO}^+)} = \frac{1}{3} \frac{k_{\rm D} ~ n({\rm HD})}{k_{\rm CO} n({\rm CO})+k_e^{\prime} n(e^-)}
\end{equation}
which can be rewritten as:
\begin{equation}
\label{eq:xe}
x_e \approx 10^{-7} \left[ \frac{0.26}{R_{\rm D}} - \frac{10}{f_d} \right]
\end{equation}
where the gas temperature is $T = 10$~K, the CO fractional abundance $x({\rm CO}) \sim 9.5 \times10^{-5}$, and the abundance of deuterium relative to hydrogen $\sim 1.5 \times 10^{-5}$. $f_d$ is the fraction of CO in gaseous form (not condensed into grains). Numerical values for the reaction rates are $k_{\rm D} \sim 1.7 \times 10^{-9}$~cm$^3$/s, $k_{\rm CO} \sim 6.6 \times 10^{-10} (T/300~{\rm K})^{-0.5}$~cm$^3$/s, and $k_e^{\prime} \sim 6.0 \times 10^{-8} (T/300~{\rm K})^{-0.5}$~cm$^3$/s  \cite{paola,ceccadominik}.
It follows that values of $R_{\rm D}$ larger than $\sim 0.026 f_d$ cannot be accounted for by this simple model \cite{caselli}.
Moreover, a way of reasoning very close to that used to determine the expression of $R_{\rm D}$ leads to:
\begin{equation}
R_{\rm H} = \frac{n({\rm HCO}^+)}{n({\rm CO})} = \frac{k_{\rm CO}[\zeta_{\rm CR}^{{\rm H}_2}/n({\rm H}_2)]}{\beta_e x_e [k_{\rm CO} x({\rm CO})+k_e x_e]}
\end{equation}
where $\beta_e = 2 \times 10^{-7} (T/300 {\rm K})^{-0.75}$~cm$^3$/s and $k_e = 6.8 \times 10^{-8} (T/300 {\rm K})^{-0.5}$~cm$^3$/s represents the reaction rate for the dissociative recombination of HCO$^+$ and H$_3^+$, respectively. 
Hence, an expression for the CR ionization rate:
\begin{equation}
\label{eq:zeta}
\frac{\zeta_{\rm CR}^{{\rm H}_2}}{n({\rm H}_2)} \approx \left( 2.6 \times 10^{-4} x_e + \frac{2.4 \times 10^{-10}}{f_{\rm D}} \right) x_e R_{\rm H}
\end{equation}
which has been written in a way to highlight the fact that the ratio between the CR ionization rate and the H$_2$ density, rather than the two quantities separately, plays a key role in cloud chemical models.
Even though Eqns.~\ref{eq:xe} and \ref{eq:zeta} stem from an oversimplified description of the cloud chemistry (see \cite{paola} for a discussion), they are still illustrative of the fact that a measurement of the abundance ratios $R_{\rm D}$ and $R_{\rm H}$ gives an estimate of the electron fraction $x_e$ and CR ionization rate.

Before proceeding to apply the methods described above to specific astrophysical objects, it is worth reminding that the study of other molecular lines (e.g. H$_2^+$, OH$^+$, H$_2$O$^+$, H$_3$O$^+$) has been proposed/used in order to infer the value of the CR ionization rate in MCs (e.g. \cite{alternative}). 

\section{Low and high energy observations of SNR/MC associations: W28, W51C, IC443}
\label{sec:SNRMC}

Fig.~\ref{fig:w28} displays the multi-wavelength emission from the region surrounding the aged (few $10^4$ yr) SNR W28, whose boundary is approximated by the blue circle. A MC complex is present in the region, as revealed by the CO emission (red and magenta contours). The northern part of the MC complex is definitely associated with the SNR shell, as indicated by the presence of OH masers, unambiguous sign of shock/cloud interaction \cite{hessw28}. The association between the SNR and the southern part of the MC complex is less clear, due to the large errors in the determination of its distance, which could be as distant as $\approx 4$~kpc, while the estimated distance of the SNR is $\approx 2$~kpc \cite{hessw28}. In the following we will assume the whole complex to be associated to the SNR and located at a distance of $\approx 2$~kpc, and will explore the implication of such configuration.

Two very high energy gamma-ray sources have been reported in the W28 region by the HESS collaboration (greyscale and white contours in Fig.~\ref{fig:w28}), HESS J1801-233 and HESS J1800-240, the latter structured into three components indicated by the letters A, B, and C. 
The first remarkable fact to be noted is the good spatial correlation between CO and TeV emission. In particular, the locations of the gamma-ray sources HESS J1801-233 and HESS J1800-240 A and B correspond to the three most prominent peaks in the CO emission.
Such spatial correlation is suggestive of a scenario in which the gamma rays are produced in CR hadronic interactions with the dense gas of the MC complex (see Sec.~\ref{sec:gamma}). 
The masses of the northern and southern components of the MC complex, coinciding with the gamma-ray sources HESS J1801-233 and HESS J1800-240 have been estimated to be $5 \times 10^4$ and $10^5 M_{\odot}$, respectively \cite{hessw28}.
The very high energy (TeV) gamma-ray spectra of the gamma-ray sources HESS J1801-233 and HESS J1800-240 have been fitted with power laws of the form $F_{\gamma}(E_{\gamma}) = A \times E_{\gamma}^{-\Gamma}$ with best fit values equal to $A = 7.50$ and $18.63 \times 10^{-13}$ TeV$^{-1}$ cm$^{-2}$ s$^{-1}$ for the spectral normalization and $\Gamma = 2.66$ and $2.49$ for the spectral slope, respectively. 
We can now use Eq.~\ref{eq:gammarayflux} to estimate the excess $\delta$ of multi-TeV CRs present in the clouds. This gives $\delta = 23$ and $42$, for the northern and southern cloud, respectively.
Thus, MCs in the W28 region are bombarded by multi-TeV CRs whose intensity exceeds by a factor of several tens that of the sea of Galactic CRs.
This fact has been interpreted as an indication that the SNR W28 is (or most likely was, in the past) a powerful accelerator of CRs up to energies exceeding the TeV, and that the CRs escaped the SNR and now fill an extended region around it \cite{gabiciw28}.

\begin{figure}[t]
\centering
\includegraphics[width=0.51\textwidth,clip]{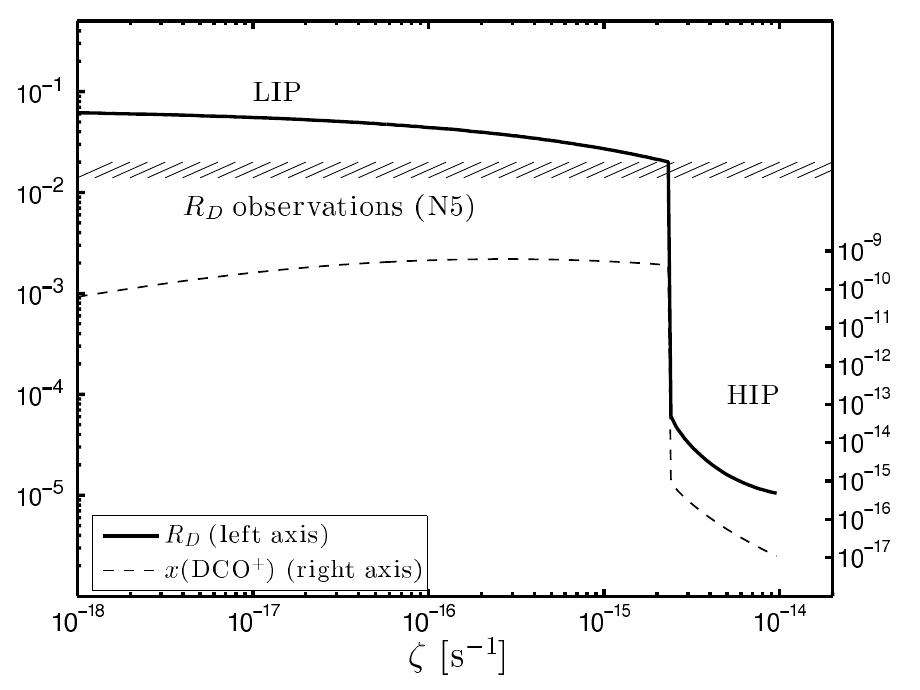}
\includegraphics[width=0.48\textwidth,clip]{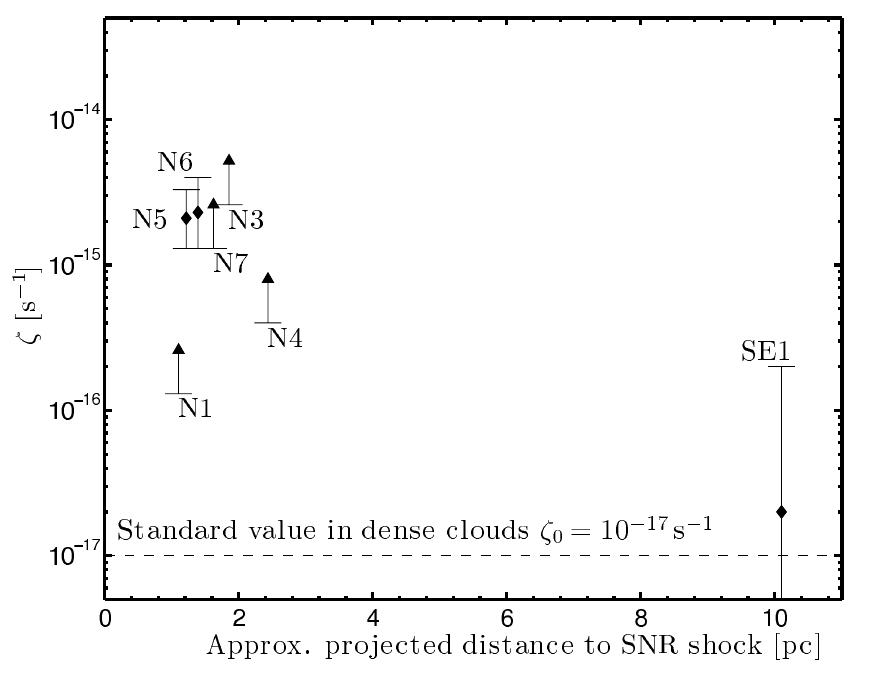}
\caption{{\bf Left:}  $R_{\rm D} = {\rm DCO}^+/{\rm  HCO}^+$ (thick line, left axis) and $x({\rm DCO}^+) = n({\rm DCO}^+) /n_{\rm H}$ (dashed line, right axis) as a function of $\zeta_{\rm CR}^{{\rm H}_2}$, for T = 10 K and $n_{\rm H} = 8 \times 10^3$ cm$^{-3}$, i.e. the physical conditions characterizing position N5 (see green crosses in Fig.~1). The high-ionization (HIP) and low-ionization (LIP) phases are marked. The hatched area shows the range of observed $R_{\rm D}$ at position N5. {\bf Right:} CR ionization rate $\zeta_{\rm CR}^{{\rm H}_2}$ as a function of the projected distance from the SNR shock (blue circle in Fig.~1), assuming a 2 kpc distance.}
\label{fig:results}       
\end{figure}

The MCs in the W28 region have been also targeted by the IRAM millimeter telescope (green crosses in Fig.~\ref{fig:w28}) to search for emission lines from HCO$^+$ and DCO$^+$ \cite{w28noi}.
Of the 16 positions observed, 10 are from the northern cloud, and 6 from the southern one. The abundance ratio $n({\rm DCO}^+)/n({\rm HCO}^+)$ ratio  and the CR ionization rate was determined for three positions (two in the north and one in the south), and upper limits on the abundance ratio (corresponding to lower limits on the CR ionization rate) were obtained for four positions, all in the northern cloud.

In fact, Vaupré et al. \cite{w28noi} did not use the analytic method described in Sec.~\ref{sec:chemistry}, but a more sophisticated numerical model accounting for a much larger chemical network. Results from these computations show that, as expected, the ionization fraction $x_e$ increases together with $\zeta_{\rm CR}^{{\rm H}_2}/n({\rm H}_2)$, while $R_{\rm D}$ decreases. 
For values of $R_{\rm D}$ larger than $\approx 2 \times 10^{-2}$ the agreement between the analytic and the numerical model is very good, while for smaller values the difference becomes dramatic.
This is because when $R_{\rm D}$ decreases below $\approx 2 \times 10^{-2}$ the numerical model predicts an abrupt jump to a regime characterized by large values of $x_e$ and very low values of $R_{\rm D}$.

This jump, shown in Fig.~\ref{fig:results} (left panel), corresponds to the well-known transition from the so-called low ionization phase (LIP) to the high ionization phase (HIP), and is due to the sensitivity of interstellar chemical networks to ionization (see \cite{w28noi} and references therein). 
Though a detailed analysis of the transition between the LIP and the HIP phase is clearly beyond the scope of this paper, the existence itself of this instability is of great interest since it produces a sharp difference between the analytical and the numerical predictions from the ionization point of view. 
While in the analytic model described in Sec.~\ref{sec:chemistry} the variations of $x_e$ and $R_{\rm D}$ are continuous and predict a scaling $x_e \sim 1/R_{\rm D}$ for low values of $R_{\rm D}$ (Eq.~\ref{eq:xe}), in the numerical models both $x_e$ and $R_{\rm D}$ undergo a sudden jump at a given value of $\zeta_{\rm CR}^{{\rm H}_2}/n({\rm H}_2)$.
To summarize: the LIP is associated with $R_{\rm D}$ larger than $\approx 10^{-2}$ and $x(e^-) \lesssim 5 \times 10^{-7}$, whilst the HIP is characterized by values of both $R_{\rm D}$ and $x(e^-)$ of the order of few times  $10^{-5}$. In our calculations, the LIP-HIP transition occurs at $\zeta_{\rm CR}^{{\rm H}_2}/n({\rm H}_2) \sim 3 \times 10^{-19}$~cm$^3$/s, with a quite weak dependence on the gas temperature.

As shown in Fig.~\ref{fig:results} (left panel), the low values of $R_{\rm D}$ in the HIP are due to a very low (virtually undetectable) abundance of DCO$^+$, whilst the abundance of HCO$^+$ decreases by a smaller amount. This has important consequences when using the DCO$^+$/HCO$^+$ method to derive the ionization fraction and CR ionization rate. First, it must be recognized that this method may provide a value of $x_e$ only for LIP-dominated gas conditions. In other words, where DCO$^+$ is detected, the line of sight is dominated by low-$x_e$ gas. For lines of sight dominated by HIP gas, the abundance of DCO$^+$ is expected to be well below detectable thresholds, such that only upper limits on $R_{\rm D}$, and thus lower limits on the CR ionization rate $\zeta_{\rm CR}^{{\rm H}_2}$, can be derived. 

The dependence of $\zeta_{\rm CR}^{{\rm H}_2}$ on the projected distance from the radio boundary of W28 is shown in Fig.~\ref{fig:results} (right panel). Remarkably, the point farthest ($\approx$~10 pc) from the SNR edge (point SE1, Fig.~\ref{fig:w28}) is the one with the lowest value of $\zeta_{\rm CR}^{{\rm H}_2}$. Actually, it is the only point where the gas is predominantly in the LIP state and the CR ionization rate is comparable to the Spitzer value. All other points, at distances $\lesssim$~3 pc from the SNR shock, have at least a fraction of the gas in the HIP, namely they have a larger $x_e$, and $\zeta_{\rm CR}^{{\rm H}_2} \gtrsim 10^{-15}$~s$^{-1}$. 
We can conclude, therefore, that the MC next to the SNR is irradiated by an enhanced flux of CRs of relatively low energy, while the cloud farther away is not.

\begin{figure}[t]
\centering
\includegraphics[width=0.32\textwidth,clip]{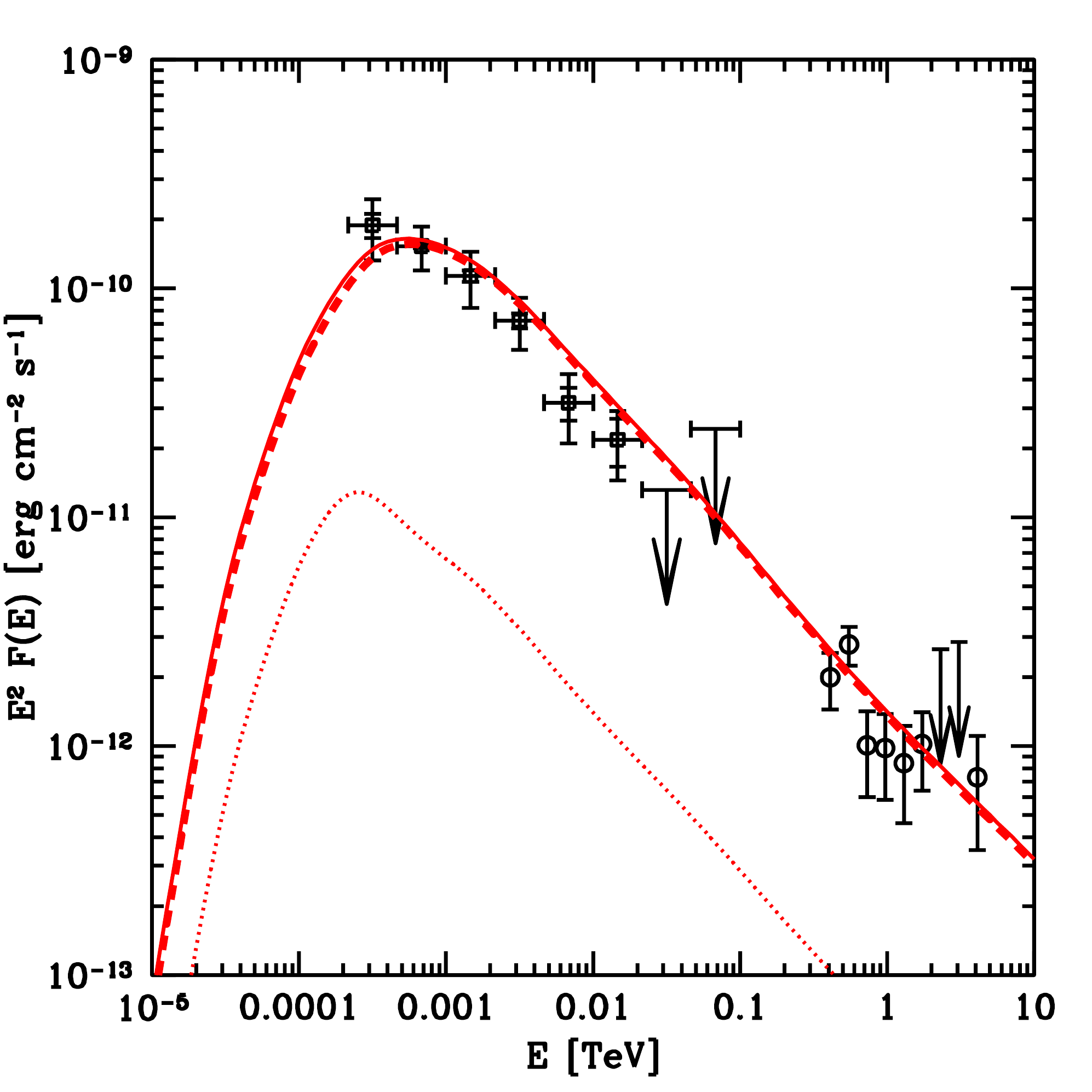}
\includegraphics[width=0.32\textwidth,clip]{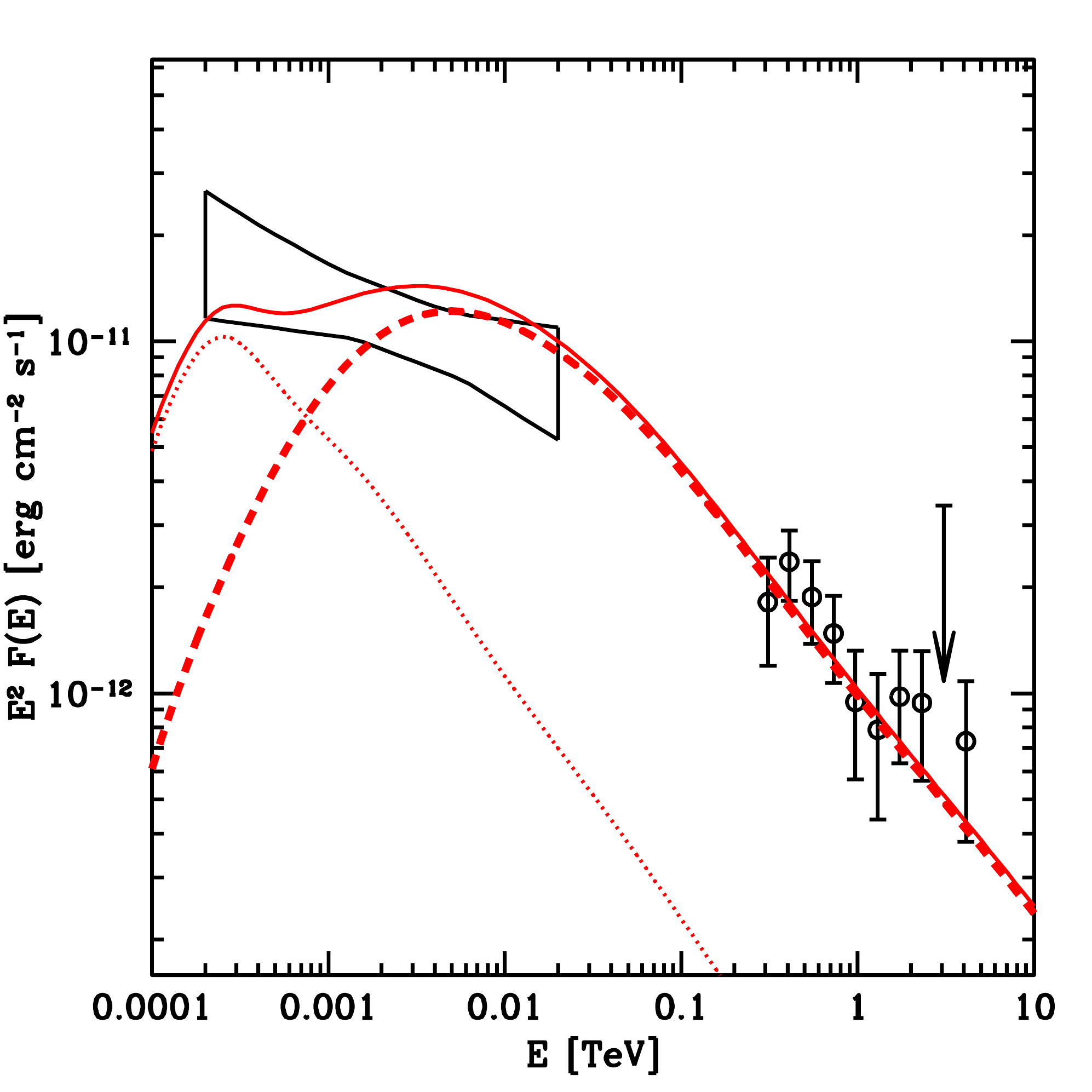}
\includegraphics[width=0.32\textwidth,clip]{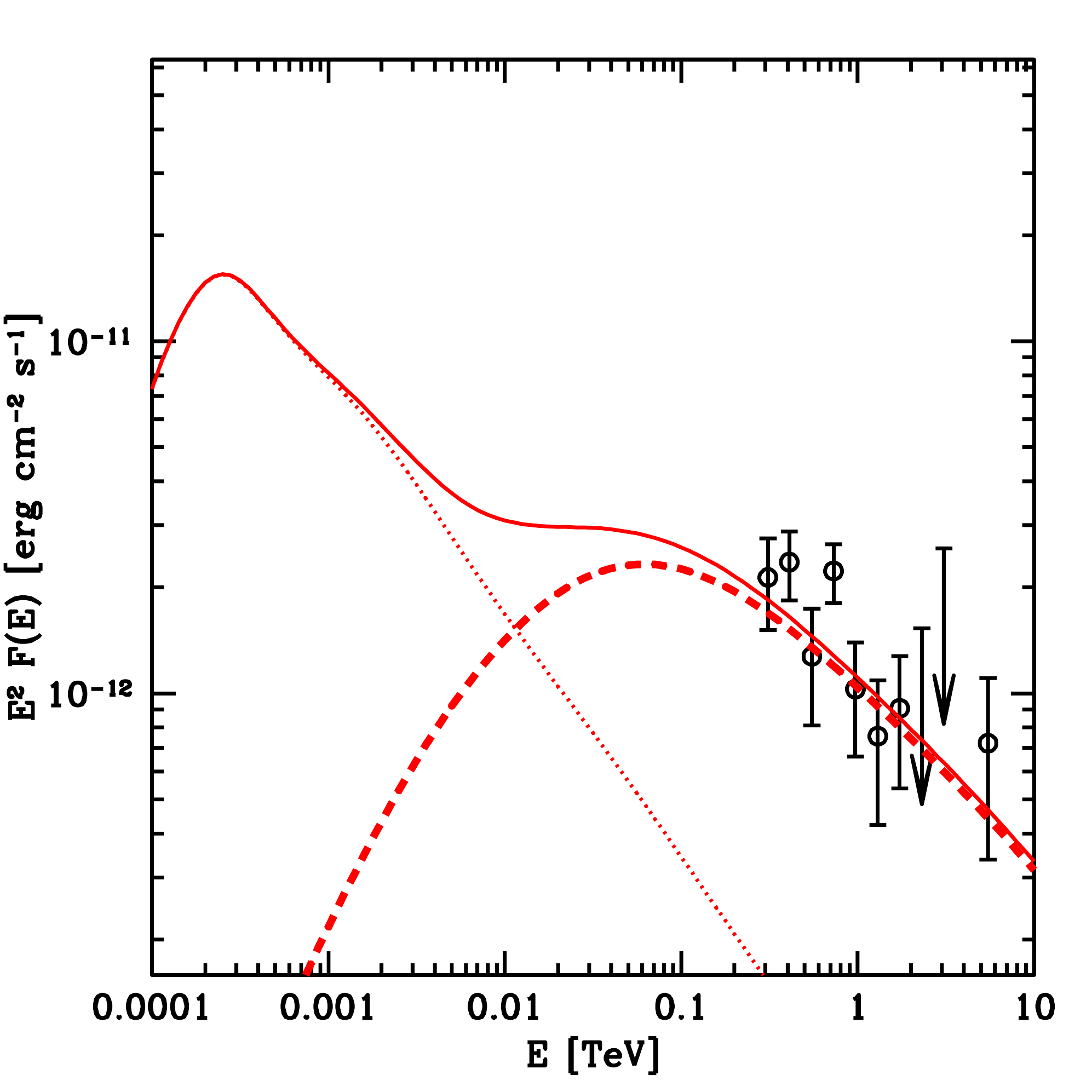}
\caption{Fermi and HESS data points for HESS J1801-233, HESS J1800-240 B and A (left to right), the latter not detected by Fermi. Dotted lines represent the expected emission from the CR sea (subtracted as a background in the Fermi data analysis) while dashed lines the emission from CRs escaped from W28 \cite{gabiciw28}. The fit is obtained for a distance from the SNR that increases gradually from the left to the right panel.}
\label{fig:sf2a}       
\end{figure}

We can now combine the constraints coming from the HCO$^+$/DCO$^+$ method with those coming from gamma-ray observations. Fig.~\ref{fig:sf2a} shows the gamma-ray emission from the sources HESS J1801-233, HESS J1800-240 B, and HESS J1800-240 A (left to right), which coincide with the three prominent peaks in the CO emission. Data points are from Fermi (GeV domain) and HESS (TeV domain). 
Both the northern (HESS J1801-233) and southern clouds (HESS J1800-240 A and B) coincide with sources of TeV emission, as seen by HESS. This means that the clouds are illuminated by very high energy ($\approx$~10 TeV) CRs, which already escaped the SNR shell and travelled the $\gtrsim$~10 pc (or more, if projection effects play a role) to the southern cloud. Conversely, the low CR ionization rate measured in SE1 tells us that the ionizing lower energy CRs remain confined closer to the SNR. In the same vein, GeV emission has been detected towards the northern region but only towards a part of the southern one (from HESS J1800-240 B only, see Fig.~\ref{fig:sf2a}). This difference between the GeV and TeV gamma-ray morphology has been interpreted as a projection effect: the portion of the southern region that exhibits a lack of GeV emission is probably located at a distance from the shock significantly larger than the projected one, >10 pc, and thus can be reached by $\gtrsim$~TeV CRs but not by $\gtrsim$~GeV ones \cite{gabiciw28}. Remarkably, the SE1 point is located in the region where the lack of GeV emission is observed. 
The gamma-ray emission from the MC complex has been successfully fitted within this framework (red lines in Fig.~\ref{fig:sf2a} \cite{gabiciw28}).

The picture that emerges is that of a stratified structure with CRs of larger and larger energies occupying larger and larger volumes ahead of the shock. Within this framework, it is possible to estimate the CR diffusion coefficient in the region. This can be done by recalling that in a given time $t$, CRs diffuse over a distance $R \approx \sqrt{D \times t}$, where D is the energy dependent diffusion coefficient. Thus, the detection of GeV emission from the southern cloud requires:
\begin{equation}
\label{eq:diffusion}
D(10~ {\rm GeV}) \gtrsim 3 \times 10^{27} \left( \frac{R}{10~{\rm pc}} \right)^2 \left( \frac{t}{10^4~{\rm yr}} \right)^{-1} {\rm cm^2/s}
\end{equation}
where $D(10~{\rm GeV})$ is the diffusion coefficient of 10 GeV CRs, which are those responsible for the $\approx$~GeV gamma-ray emission, and $t$ is the time elapsed since the escape of CRs from the SNR. 

The diffusion coefficient obtained in Eq.~\ref{eq:diffusion} can then be rescaled to lower energies, according to $D \propto \beta p^s$, where $p$ is the particle momentum, $\beta = v/c$ its velocity in units of the speed of light, and $s$ depends on the spectrum of the ambient magnetic turbulence. The typical value of $s$ in the interstellar medium is poorly constrained to be in the range 0.3 to 0.7 \cite{jones}, and in the following we adopt a reference value $s = 0.5$. To estimate the diffusion length of low energy CRs, one has to keep in mind that, while CRs with energies above $\approx$~GeV are virtually free from energy losses (the energy loss time for proton-proton interactions in a density $n_{\rm H} \approx 10^3$~cm$^{-3}$ is comparable to the age of the SNR), lower energy CRs suffer severe ionization losses over a short timescale of the order of $\tau_{ion} \approx 14 ~ (n_{\rm H}/10^3~{\rm cm}^{-3})^{-1} (E/{\rm MeV})^{3/2}$~yr (valid in the range of energies spanning 1-100 MeV) \cite{w28noi}. 
The diffusion length of low energy CRs can now be estimated by equating the diffusion time $\tau_d \approx R_d^2/D$ to the energy loss time $\tau_{ion}$, which gives $R_d \approx$ 0.02, 0.3, and 3 pc for CRs of energy 1, 10, and 100 MeV, respectively. This implies that only CRs with energies $\gtrsim$~100 MeV can escape the shock and spread over a distance of 3 pc or more, and thus these are the CRs that play a major role in ionizing the gas. Whether the ionization of the gas is due directly to these CRs or to the products of their interaction with the gas (namely slowed down lower energy CRs or secondary products of CR interactions) remains an open question. It is remarkable that the particle energies of ionizing CRs ($\approx$~0.1-1 GeV) also make them capable of producing sub-GeV gamma rays, given that the kinetic energy threshold for $\pi^0$ production is $\approx$~280 MeV.

Very similar conclusions have been reached in \cite{sugar}, where the gamma-ray emission from HESS J1801-233 (the northern MC, characterized by a ionization rate $\gtrsim 2 \times 10^{-15}$~s$^{-1}$) has been fitted by assuming an underlying CR proton spectrum which is a power law in momentum. A good fit is obtained for $4 \pi p^2 N(p) \propto p^{-2.8}$.
The normalization of $N(p)$ can be derived from the measurement of the gamma-ray flux of the MC and from the values of the MC mass and distance (see Eq.~\ref{eq:gammarayflux}).
Once the normalization is obtained, the expected ionization rate due to CR protons can be computed.
This quantity will depend on the extension of the CR spectrum to low energies, below 280 MeV, which
are more relevant for ionization but are not probed by the observations of $\pi^0$-decay gamma-rays. 
Thus, a {\it guaranteed} level of ionization is obtained after assuming that the CR spectrum is abruptly truncated below 280 MeV, i.e. the threshold for $\pi^0$ production.
The ionization rate obtained in this way is\footnote{The contribution from nuclei heavier than H to both gamma-ray emission and ionization is not considered. Nevertheless, results are reliable, since the inclusion of heavy nuclei would lead to a comparable increase of both quantities.} $\approx 4 \times 10^{-16}$~s$^{-1}$, roughly 20\% the observed value, which implies that the CRs responsible for the gamma-ray emission give a non negligible contribution to the ionization rate.
To match the observed value $\zeta_{\rm CR}^{{\rm H}_2} \approx 2 \times 10^{-15}$~s$^{-1}$ the CR spectrum has to be extrapolated (assuming the same slope in momentum)  down to energies of $\approx$~60 MeV.
This implies that in order to be consistent with the measured ionization rate, the CR proton spectrum
cannot continue with the same slope below particle energies of $\approx$ 50-100 MeV. A break or a cutoff must be present in the CR spectrum. Also, it seems that there is not much room left for CR electron induced ionization of the gas, unless the CR proton spectrum is cut off abruptly at a particle energy just below or of the order of the threshold for neutral pion production\footnote{The CR ionization rates and the gamma-ray spectra presented in \cite{sugar} were computed by the newly released CRIME code (Cosmic Ray Interactions in Molecular Environemnts), publicly available at http://crime.in2p3.fr \cite{julian}.}.

All these facts seem to indicate that {\it a link between low and high energy CRs has been, finally, established. 
In the scenario described above, the very same CRs are responsible for both part of the observed ionization of the gas and for the production of low energy gamma rays.} If confirmed by observations of other SNR/MC associations, such a link would constitute a robust evidence for the presence of accelerated protons, of low and high energy, in the environment of SNRs, a thing that would bring support to the idea that SNR are the sources of Galactic CRs of all energies. 

A similar enhancement in the CR ionization rate was reported from a massive MC ($\approx 1.9 \times 10^5 M_{\odot}$) interacting with the aged ($\approx 3 \times 10^4$~yr) and distant ($\approx 5.5$~kpc) SNR W51C \cite{magicW51}, with a value of $\zeta_{\rm CR}^{{\rm H}_2} \approx 1.0 - 1.3 \times 10^{-15}$~s$^{-1}$ measured from DCO$^+$/HCO$^+$ observations by the IRAM 30 m telescope (red point, Fig.~\ref{fig:zeta}) \cite{ceccaW51c}.
The TeV emission from the MC, as detected by MAGIC, was fitted by the power law $F(E_{\gamma}) = A \times E_{\gamma}^{-\Gamma}$ with $A = 9.7 \times 10^{-13}$~cm$^{-2}$ s$^{-1}$ TeV$^{-1}$ and $\Gamma = 2.58$, calling for a large overdensity of multi-TeV CRs in the MC of the order of $\delta \sim 74$ (see Eq.~\ref{eq:gammarayflux}). However, it has to be stressed that part of the gamma-ray emission might indeed come form a known pulsar wind nebula candidate, and in this case the value of $\delta$ would be reduced correspondingly. On the other hand, probably only a fraction (possibly $\approx$ 0.1) of the MC mass is actually irradiated by CRs, and this would increase the value of $\delta$ \cite{magicW51}.
Follow up observations of the region characterized by the enhanced CR ionization rate led to the detection of SiO emission which typically traces slow shocks. Such shocks are naturally produced when a fast SNR blastwave engulfs dense gas clumps, indicating that the enhanced CR ionization is probably located downstream of the SNR shock \cite{dumas}.

Finally, the aged ($\approx 3 \times 10^4$~yr) and nearby ($\approx 1.5$~kpc) SNR IC443 interacting with a diffuse MC (total mass in the region $\approx 10^4 M_{\odot}$ \cite{diego}) was observed in the infrared to search for H$_3^+$ absorption lines \cite{indrioloIC443}. H$_3^+$ was detected from two of the six lines of sight observed, corresponding to a CR ionization rate of 1.6 and $2.6 \times 10^{-15}$~s$^{-1}$ (green points, Fig.~\ref{fig:zeta}). 
The non detection of H$_3^+$ from four of the lines of sight under examination has been interpreted as a possible outcome of CR propagation in the region \cite{indrioloIC443}.
Also in this case, the TeV spectrum, as detected by VERITAS, is a power law $F(E_{\gamma}) = A \times E_{\gamma}^{-\Gamma}$ with $A = 8.38 \times 10^{-13}$~cm$^{-2}$ s$^{-1}$ TeV$^{-1}$ and $\Gamma = 2.99$ \cite{veritasIC443}, implying an overdensity of multi-TeV CRs of $\delta \approx 52$.
Remarkably, the Fermi collaboration reported on the detection of the characteristic pion bump in the GeV spectrum of IC443, ascribing unambiguously the origin of the gamma-ray emission to hadronic processes \cite{pionbump}.

\section{Conclusions and future perspectives}
\label{sec:conclusions}

So far, only 3 gamma-ray bright SNR/MC associations (or SNOBs) have been observed in the infrared or millimeter domain: W28, W51C, and IC443. 
In all cases, an overdensity of multi-TeV CRs of a factor of several tens has been reported, as well as a CR ionization rate at the level of $\gtrsim 10^{-15}$~s$^{-1}$, pointing towards an excess of low energy (plausibly MeV-domain) CRs.

While the procedure to determine the CR overdensities from gamma-ray data is solid and straightforward (as long as one knows the mass of the MC under examination), those based on low energy observations suffer limitations. The observations of H$_3^+$ lines rely on the presence of a bright background infrared star, a fact which limits both the number of possible lines of sight and the MC column densities that can be probed (if the cloud is too thick the star cannot be seen).
On the other hand, the DCO$^+$/HCO$^+$ method can provide an estimate of $\zeta_{\rm CR}^{{\rm H}_2}$ only if the gas is predominantly in the LIP phase, while giving only lower limits on it if the gas is in the HIP state. 

Another limitation of the approach is connected to the fact that $\zeta_{\rm CR}^{{\rm H}_2}$ is an integral quantity (integral over particle energy of the product between CR spectrum and ionization cross section), which means that no direct information can be extracted on the CR spectrum in the low energy domain, as it is done in the high energy domain from gamma-ray data.
Thus, any extraction of the low energy CR spectrum from data has to be done through the theoretical modeling of the acceleration of CRs at SNR shocks and their propagation into MCs.
Besides the (quite large) uncertainties introduced by modeling, one has to keep in mind that we do not even fully understand why the ionization rate is different in dense and diffuse {\it isolated} MCs!
This is most likely related to the way in which CRs penetrate MCs (see e.g. \cite{giovanni}), but it is clear that, to date, a proper description of the penetration mechanism is still lacking.

Thus, in order to proceed along this line of research and reinforce the link between low and high energy CRs established by the observations reviewed in this paper, efforts should be aimed at:
\begin{itemize}
\item{increasing the number of SNR/MC associations detected at both low and high energy (SNR W44 is a quite obvious first candidate for that, given the fact that both the presence of the pion bump \cite{pionbump} and of runaway cosmic rays \cite{fermiW44bis} have been revealed by Fermi);}
\item{searching for chemical tracers of low energy CRs in MCs, alternative to HCO$^+$/DCO$^+$ and H$_3^+$, and}
\item{improving theoretical models to describe the acceleration of CRs at shocks and their propagation into MCs.}
\end{itemize}

\end{document}